\documentclass[conference]{IEEEtran}

\usepackage{cite}
\ifCLASSINFOpdf
   \usepackage[pdftex]{graphicx}
   \graphicspath{Figures/}
\else
 \fi
\usepackage[cmex10]{amsmath}
\usepackage{algorithm}
\usepackage{algpseudocode}
\usepackage{array}
\usepackage{bm}   
\usepackage{multirow} 
\usepackage{bigdelim}
 
\usepackage{booktabs}
\usepackage{color}
\def\NoNumber#1{{\def\alglinenumber##1{}\State #1}\addtocounter{ALG@line}{-1}}
\usepackage{mathtools}
\DeclarePairedDelimiter\ceil{\lceil}{\rceil}

\ifCLASSOPTIONcompsoc
  \usepackage[caption=false,font=normalsize,labelfont=sf,textfont=sf]{subfig}
\else
  \usepackage[caption=false,font=footnotesize]{subfig}
\fi

\usepackage{stfloats}
\begin{document}
%
\title{Impact of Communication Delay on Asynchronous Distributed Optimal Power Flow Using ADMM}

\author{\IEEEauthorblockN{Junyao Guo\IEEEauthorrefmark{1},
Gabriela Hug\IEEEauthorrefmark{2}, and
Ozan Tonguz\IEEEauthorrefmark{1}}
\IEEEauthorblockA{\IEEEauthorrefmark{1}Department of Electrical and Computer Engineering, Carnegie Mellon University, Pittsburgh, PA, USA\\
\IEEEauthorrefmark{2}Power Systems Laboratory, ETH Z{\"u}rich, Z{\"u}rich, Switzerland\\
Email: junyaog@andrew.cmu.edu, ghug@ethz.ch, tonguz@ece.cmu.edu}}
\maketitle



\maketitle

\begin{abstract}
Distributed optimization has attracted lots of attention in the operation of power systems in recent years, where a large area is decomposed into smaller control regions each solving a local optimization problem with periodic information exchange with neighboring regions. However, most distributed optimization methods are iterative and require synchronization of all regions at each iteration, which is hard to achieve without a centralized coordinator and might lead to under-utilization of computation resources due to the heterogeneity of the regions. To address such limitations of synchronous schemes, this paper investigates the applicability of asynchronous distributed optimization methods to power system optimization. Particularly, we focus on solving the AC Optimal Power Flow problem and propose an algorithmic framework based on the Alternating Direction Method of Multipliers (ADMM) method that allows the regions to perform local updates with information received from a subset of but not all neighbors. Through experimental studies, we demonstrate that the convergence performance of the proposed asynchronous scheme is dependent on the communication delay of passing messages among the regions. Under mild communication delays, the proposed scheme can achieve comparable or even faster convergence compared with its synchronous counterpart, which can be used as a good alternative to centralized or synchronous distributed optimization approaches.
\end{abstract}


\IEEEpeerreviewmaketitle

\section{Introduction}
\label{intro}
Optimizing the operation of the electric power systems has become an increasingly challenging task due to the increasing number of control variables associated with distributed generation and flexible loads and the system's inherently non-linear physical characteristics. Conventionally, the optimal management of a large-scale power system is solved as a single optimization problem at a centralized location, which limits the size of the problem that can be solved in a given amount of time. Moreover, as all the measurements need to be collected by a central controller, the communication load of the backbone network is high which may result in large communication delays or failures of data delivery. To address these issues, there has been a growing interest in distributed optimization, where the optimization problem associated with a large region is decomposed into subproblems each associated with a smaller region. These regions are connected via transmission lines and therefore need to communicate periodically to achieve an overall optimal solution for operating the entire grid. 

Many iterative distributed optimization methods have been proposed \cite{conejo2006decomposition}\cite{bertsekas1989parallel}, and for most of these methods, synchronization of the subproblems is required; i.e., at each iteration, all subproblems need to be solved and then start next iteration. However, synchronization may not be easily acquired in a distributed system without a centralized coordinator because a region may not know how long to wait for other regions a priori before proceeding to its next iteration. Furthermore, even if synchronization is achievable, it may lead to an inefficient implementation of distributed methods. The sizes and complexities of the control regions are dependent on the power system's physical configuration. Therefore, these regions are usually heterogeneous and require different amounts of computation time. Moreover, the communication delays among these regions are also heterogeneous which are dependent on the communication infrastructures and network topologies deployed. In a synchronous scheme, as all regions need to wait for the slowest region to finish its computation or data transmission, some regions would remain idle for most of the time which results in under-utilization of both the computation and communication resources.

The synchronization issue has been systematically studied in the research fields of distributed computing with seminal works \cite{bertsekas1989parallel}\cite{dwork1988consensus}. While the concept of asynchronous iterative computing is not new, it remains an open question whether those methods can be applied to complex systems such as power systems \color{black}which have non-convex physical characteristics. Most existing asynchronous distributed optimization methods can only tackle convex optimization problems \cite{bertsekas1989parallel, dwork1988consensus, peng2016arock}, and therefore can be only applied to solve approximations of the non-convex problems arising in power systems \cite{7285913}\cite{abboud2014asynchronous}\cite{nguyen2016distributed}. Several asynchronous algorithms that tackle problems with some level of non-convexity are recently proposed with applications in the wireless sensor network \cite{kumar2017asynchronous} and machine learning \cite{7423789}. But the problem considered in these studies is the consensus problem where the nodes are usually homogeneous and local updates are easy to compute. However, many distributed optimization applications in power system entail a different problem formulation from the consensus problem where geographical partitioning of the system is usually considered and neighboring control regions are coupled by non-convex AC power flow constraints, which are not handled in \cite{kumar2017asynchronous}\cite{7423789}. Moreover, different from the schemes in \cite{7285913}\cite{7423789} that require a master node and are designed for a parallel computing environment with a cluster of computers, the computational entities for distributed optimization in power systems are placed at distant locations and the existence of any centralized node might harm the scalability of the algorithm. \color{black}


To provide insight into the applicability of asynchronous distributed methods to power system optimization, in this paper, we propose an algorithmic framework that allows each region to perform local updates in an asynchronous fashion \color{black} without any centralized coordination\color{black}. The optimization problem studied is the non-convex AC Optimal Power Flow (AC OPF) problem. The proposed framework assumes a message-passing model where each region is allowed to solve its local OPF problem with partial but not all updated information received from its neighbors. Our proposed algorithm is based on the ADMM method used in \cite{erseghe2015distributed} and we extend this method to fit into the asynchronous framework with convergence analysis provided. Particularly we study the impact of communication delay on the convergence performance of the proposed method, whereas the aforementioned studies have not investigated the role of communications in their proposed asynchronous methods. Through experimental studies, we show that under mild communication delays, the proposed asynchronous ADMM approach could reduce the execution time compared with its synchronous counterpart by reducing the waiting time for slow regions. However, the performance of asynchronous ADMM deteriorates under large communication delays. These findings indicate that asynchronous distributed computing schemes could be beneficial for the operation of power systems, with the premise that the communication infrastructure or schemes are carefully chosen to support relatively fast communications.

The rest of the paper is organized as follows: Section \ref{ACOPF} formulates the AC OPF problem. Section \ref{synchronousADMM} presents the synchronous distributed ADMM and its application to solving the AC OPF problem. Section \ref{asynchronousADMM} proposes an asynchronous distributed ADMM approach based on a message-passing model that is well-suited for the considered partial mesh network. Simulation results are given in Section \ref{simulation} where we demonstrate the impact of communication delays on the level of asynchronism of the system, the choice of algorithm parameter, and the convergence speed and solution quality of the proposed asynchronous approach. Finally, Section \ref{conclusion} concludes the paper and proposes possible future directions.

\section{Problem Formulation}
\label{ACOPF}
We consider the standard AC OPF problem, where the objective is to minimize the total generation cost. The OPF problem is formulated as follows:
\allowdisplaybreaks
\begin{subequations}
\label{centOPF}
\begin{align}
\label{eq1}
\underset{V, P, Q} {\text{minimize}}~~&f({P})=\sum_{i=1}^{n_{b}} \left(a_{i}P_{i}^2+b_{i}P_{i}+c_{i}\right) \\
\label{pf}
\text{subject to}~~&P_{i}+jQ_{i}-P_{i}^{\text{load}}-jQ_{i}^{\text{load}}=V_i\sum\limits_{j\in {{\Omega }_{i}}}Y_{ij}^*V_j^*\\
\label{eqPlimit}
&P_{i}^{\min}\leq P_{i}\leq P_{i}^{\max}\\
\label{eqQlimit}
&Q_{i}^{\min}\leq Q_{i}\leq Q_{i}^{\max}\\
\label{eqVlimit}
&V_{i}^{\min}\leq |V_{i}|\leq V_{i}^{\max},
\end{align}
\end{subequations}
for $i=1, \ldots, n_{b}$ where $n_{b}$ is the number of buses. Here, $(V_{i}, P_{i}, Q_{i})$ denote the complex voltage, the active power generation and the reactive power generation at bus $i$.$(a_i,b_i,c_i)$ are the cost parameters of generator at bus $i$. $Y_{ij}$ is the $ij$-th entry of the line admittance matrix, and $\Omega_i$ is the set of buses connected to bus $i$. This problem is non-convex due to the non-convexity of the AC power flow equations (\ref{pf}). We omit the line thermal limits in this paper to keep the presentation simple, whereas in \cite{guo2016acase} the inclusion of this constraint is discussed and it has been shown that the resulting problem can also be solved by the distributed ADMM approach. 

\section{Synchronous Distributed ADMM}
\label{synchronousADMM}
Geographical decomposition of the system is considered in this paper where a power grid is partitioned into a number of smaller regions each solving a local OPF problem. In the following analysis, we use $K$, $\mathcal{T}$ and $\mathcal{N}_{k}$ to denote the total number of regions, the set of inter-region tie lines and the set of neighboring regions that connect to region $k$ via transmission lines, respectively. We also introduce $\mathcal{R}_{k}, k=1,..., K,$ to denote the set of buses included in region $k$ with $\mathcal{R}_{k}\cap\mathcal{R}_{l}=\emptyset, \forall l\neq k$. 

In Problem (\ref{centOPF}), the power flow balance constraints (\ref{pf}) at the boundary buses couple the neighboring regions which prevent them from solving local OPF problems independently. To remove such coupling, the voltages at the boundary buses of each region are duplicated. Assume region $k$ and region $l$ are connected via tie line $ij$ where $i \in {R}_{k}$ and $j \in {R}_{l}$. The voltages at bus $i$ and bus $j$ are duplicated, and the copies assigned to region $k$ are $V_{i,k}$ and $V_{j,k}$. Similarly, region $l$ is assigned the copies $V_{i,l}$ and $V_{j,l}$. To ensure equivalence with the original problem, constraints $V_{i,k}=V_{i,l}$ and $V_{j,k}=V_{j,l}$ are added to the problem. The set $\mathcal{V}_{k}$ is also introduced to denote the joint set of $\mathcal{R}_{k}$ and the duplicates of buses in $\mathcal{N}_{k}$ that are directly connected to buses in $\mathcal{R}_{k}$. 

To apply the distributed ADMM approach used in \cite{erseghe2015distributed}\cite{guo2016acase}, for each tie line $ij$, we introduce two auxiliary variables $z_{i,j}^{+}$ and $z_{i,j}^{-}$ and transform the aforementioned additional constraints into their following equivalent form 
\begin{equation}
\begin{aligned}
z_{i,j}^{-} := \beta^{-}(V_{i,k}-V_{j,k})&=\beta^{-}(V_{i,l}-V_{j,l})\\
z_{i,j}^{+} := \beta^{+}(V_{i,k}+V_{j,k})&=\beta^{+}(V_{i,l}+V_{j,l}).
\end{aligned}
\label{eq:zV}
\end{equation}
Here $\beta^{-}$ and $\beta^{+}$ are scaling factors, where $\beta^{-}$ is set to be larger than $\beta^{+}$ to emphasize on $V_{i}-V_{j}$ which is strongly related to the line flow through tie line $ij$ \cite{erseghe2015distributed}. 

By introducing \mbox{${x}_{k}=\{(V_i, P_{i}, Q_{i})~|~i \in \mathcal{V}_{k}\}$} and ${z}_{k}=\{ (z_{i,j}^{-},  z_{i,j}^{+})~|~i,j \in \mathcal{V}_{k}, (i,j) \in \mathcal{T}\}$ to denote all the primal variables and auxiliary variables in region $k$, respectively, Problem (\ref{centOPF}) can now be expressed in terms of local OPF problems:
\begin{subequations}
\label{eq:OPF}
\begin{align}
\label{diseq1}
\underset{{x}, {z}} {\text{minimize}} ~~~&\sum_{k} f_{k}({x}_{k})\\
\label{diseq2}
\text{subject to}~~~& A_{k}{x}_{k}={z}_{k},~\forall k\\
\label{diseq3}
&{x}_{k} \in \mathcal{X}_{k}, ~\forall k
\end{align}
\end{subequations}
where $f_{k}({x}_{k})$ is the local objective function of region $k$. Constraint (\ref{diseq2}) is acquired by expressing (\ref{eq:zV}) using $x_{k}$ and $z_{k}$. Constraints (\ref{diseq3}) include the local feasibility constraints (\ref{pf})-(\ref{eqQlimit}) for $\forall i \in\mathcal{R}_{k}$ and constraint (\ref{eqVlimit}) for $\forall i \in\mathcal{V}_{k}$. Let $\lambda_{k}$ denote the Lagrange multiplier associated with constraint (\ref{diseq2}) and define the following Augmented Lagrangian function
\begin{equation}
\begin{aligned}
L({x},{z},{\lambda})=&\sum_{k} \Big\{ f_{k}({x}_{k})+{\lambda}_{k}^{\top}(A_{k}{x}_{k}-{z}_{k})\\&+\frac{\rho_k}{2}\|A_{k}{x}_{k}-{z}_{k}\|^{2} \Big\},
\end{aligned}
\label{eq:Aug}
\end{equation}
where $\rho_k > 0$ is a penalty parameter which can be different for different regions. The ADMM method minimizes (\ref{eq:Aug}) by iteratively carrying out the following updating steps \cite{boyd2011distributed}:
\begin{subequations}
\label{eq:ADMMIter}
\begin{align}
\label{zupdate}
z-\text{update}: & ~~~~{z}^{\nu+1}=\text{argmin}~~L({x}^{\nu},{z},{\lambda}^{\nu})\\
\label{xupdate}
x-\text{update}: &~~~~{x}^{\nu+1} = \text{argmin}_{{x} \in \mathcal{X}}~~ L({x},{z}^{\nu+1},{\lambda}^{\nu})\\
\label{lambdaupdate}
\lambda-\text{update}: &~~~~{\lambda}^{\nu+1}={\lambda}^{\nu}+{\rho}^{\nu}(A{x}^{\nu+1}-{z}^{\nu+1}),
\end{align}
\end{subequations}
where $\nu$ denotes the counter of iterations. 
With ${z}$ fixed, each subproblem in the $x$-update only contains the local variable ${x}_{k}$, which can be solved independent of others. The $\lambda$-update can also be performed locally. The $z$-update requires the information from two neighboring regions. By minimizing (\ref{eq:Aug}), $(z_{i,j}^{-},  z_{i,j}^{+})$ for any tie line $ij, i\in \mathcal{R}_{k}, j\in \mathcal{R}_{l} $ can be calculated by
\begin{subequations}
\label{eq:zcalculate}
\begin{align}
z_{i,j}^{+} &= \frac{\lambda_{k,[ij]}+\lambda_{l,[ij]}+\rho_{k}A_{k,[ij]}x_{k} +\rho_{l}A_{l,[ij]}x_{l}}{\rho_{k}+\rho_{l}}\\
z_{i,j}^{-} &= \frac{\lambda_{k,[ij]}-\lambda_{l,[ij]}+\rho_{k}A_{k,[ij]}x_{k} -\rho_{l}A_{l,[ij]}x_{l}}{\rho_{k}+\rho_{l}}.
\end{align}
\end{subequations}
Here, the subscript $[ij]$ chooses the element in $\lambda$ and the row in $A$ that are corresponding to tie line $ij$. Thereby, region $k$ can update $z_{k}$ locally once it receives $\lambda_{l}$, $\rho_{l}$ and $A_{l}x_{l}$ from $\forall l \in \mathcal{N}_{k}$.

To ensure the convergence of the variables $x$ and $z$ to finite values for non-convex problems, the penalty $\rho$ is usually increased over the iterative process. Similar to the method used in \cite{erseghe2015distributed}, we apply the following updating rule
\begin{subequations}
\begin{align}
\label{eq:rhocompute}
\tilde{{\rho}}^{\nu+1}_k=& \left\{ \begin{array}{cl}
{\rho}^{\nu}_k  & \text{if~~} \Gamma_k^{\nu+1}\leq \gamma \Gamma_k^{\nu}\\
\tau{\rho}^{\nu}_k & \text{otherwise}
\end{array}\right. \\
\label{eq:rhoupdate}
\rho_{k}^{\nu+1}=&\max\{\tilde{\rho}_{l}^{\nu+1},\forall l \in \{k\} \cup \mathcal{N}_{k}\} \end{align}
\end{subequations}
with constants $0<\gamma<1$ and $\tau>1$. $\Gamma_k^{\nu+1}=\|A_{k}{x}_{k}^{\nu+1}-{z}_{k}^{\nu+1}\|_{\infty}$ is defined as the primal residue\cite{boyd2011distributed}. Penalty ${\rho}_{k}$ is first updated for each region via (\ref{eq:rhocompute}) after the ${\lambda}$-update and the updated penalty is exchanged between neighboring regions. Then $\rho_{k}$ is adjusted via (\ref{eq:rhoupdate}) by using the maximum penalty in the neighborhood of region $k$. 

The synchronous distributed ADMM algorithm is presented in Algorithm \ref{algADMMsync}. The stopping criterion is defined as that both the primal residue ($\Gamma_k, \forall k$) and the maximum constraint mismatch are smaller than some $\epsilon$ \cite{boyd2011distributed}\cite{guo2016acase}. Under the considered non-convex setting, the convergence of this ADMM approach to feasible $x^{*}$ and $z^{*}$ is proved in \cite{erseghe2015distributed} with the assumption that both $x$ and $\lambda$ are bounded and that a local minimum can be identified when solving the local OPF problems. Note that with an increasing penalty, \color{black}Algorithm \ref{algADMMsync} generally converges to a suboptimal solution around a local optimal point due to the fact that some difficulty may arise in finding local optimum in the $x$-update while $\rho$ is large. However, a suboptimal solution that is close to the optimal point satisfies the requirements of many practical applications.\color{black}
\begin{algorithm}[t]
\caption{Synchronous ADMM}
\label{algADMMsync}
\begin{algorithmic}[1]
\State \textbf{Initialization} Given ${x}^{0}$, set ${\lambda}^{0}={0}$, ${\rho}_{k}=\rho_{0} $, $\nu=0$ 
\State \textbf{Repeat}
\State\hspace{\algorithmicindent}Set $\nu \leftarrow \nu+1$
\State\hspace{\algorithmicindent}\textbf{Update}
\NoNumber{~~~~~Update $z$ using (\ref{eq:zcalculate})} 
\NoNumber{~~~~~Update $x$ by solving the local OPF problem}
\begin{align}
\nonumber {x}_{k}^{\nu} = & \underset{{x}_{k}\in \mathcal{X}_{k}}{\text{argmin}} ~f_{k}({x}_{k})+{\lambda}_{k}^{\nu-1 \top}(A_{k}{x}_{k}-{z}_{k}^{\nu})\\
\nonumber &+\frac{{\rho}_k^{\nu-1}}{2}\|A_{k}{x}_{k}-{z}_{k}^{\nu}\|^{2}
\end{align}
\NoNumber{~~~~~Update ${\lambda}$ using}
\begin{equation}
\nonumber{\lambda}_{k}^{\nu}={\lambda}_{k}^{\nu-1}+({\rho}^{\nu-1}_k)(A_{k}{x}_{k}^{\nu}-{z}_{k}^{\nu})
\end{equation}
\NoNumber{~~~~~Update $\tilde{{\rho}}$ according to (\ref{eq:rhocompute})}
\State\hspace{\algorithmicindent}All regions exchange the updated $Ax$, $\lambda$ and $\tilde{{\rho}}$
\State\hspace{\algorithmicindent}Update ${\rho}$ using (\ref{eq:rhoupdate})
\State \textbf{Until} a predefined stopping criterion is satisfied
\end{algorithmic}
\end{algorithm}

\section{Asynchronous Distributed ADMM}
\label{asynchronousADMM}
\begin{figure}[t]
\setlength{\abovecaptionskip}{0.2cm} 
\centering
\captionsetup[subfigure]{captionskip= 0 cm}
\subfloat[synchronous, $p=1$]
{
\label{fig:illusync}
\includegraphics[trim = 0mm 0mm 0mm 0mm, clip=true,width=7cm,height = 1.8cm]{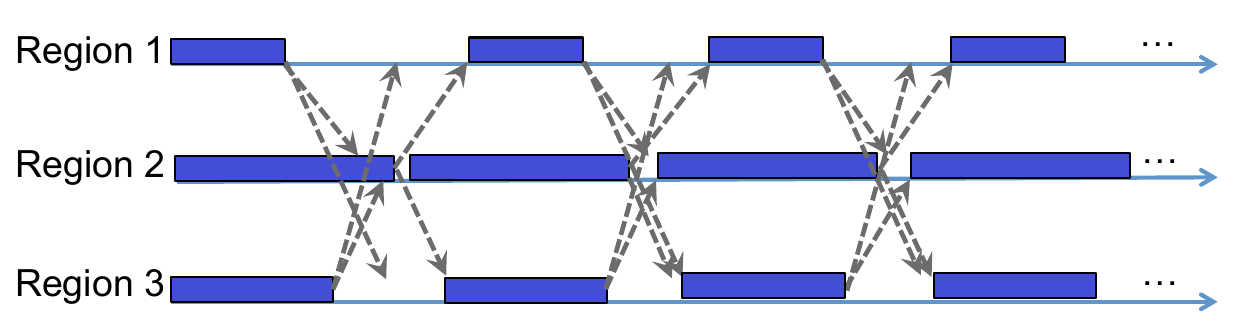} 
}
\\
\subfloat[asynchronous, $p=0.1$]
{
\label{fig:illuasync}
\includegraphics[trim =0mm 0mm 0mm 0mm, clip=true,width=7cm,height = 1.8cm]{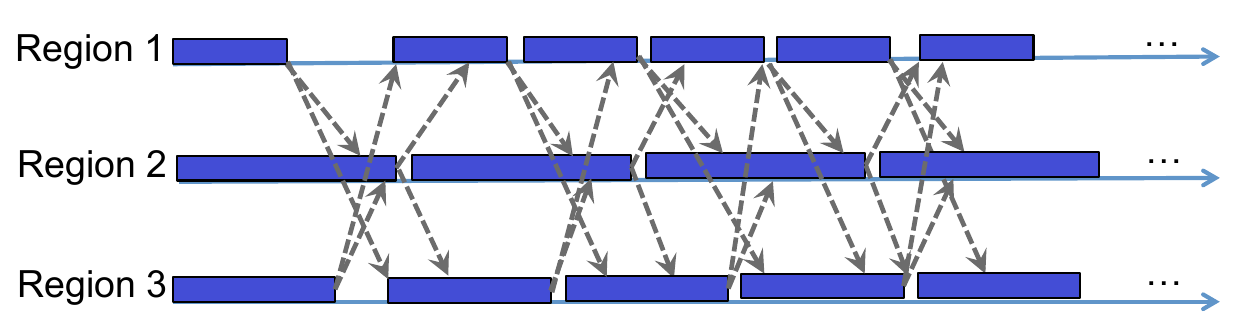} 
}
\caption{Illustration of synchronous and asynchronous distributed ADMM.}
\label{fig:asyncillustration}
\end{figure}

In this section, we propose an algorithmic framework that extends Algorithm \ref{algADMMsync} into an asynchronous setting. The proposed framework utilizes the message-passing model \cite{dwork1988consensus} where each region determines when to carry out its next iteration of local computation based on the messages it receives from its neighbors. We say that a neighbor $l$ is `arrived' at region $k$ if the updated information of $l$ is received by $k$. We assume that each region always sends out its updated information to neighbors and this information would eventually arrive at its destination; i.e., \textit{the delay is bounded}. This assumption guarantees that each region could get the information from all of its neighbors once in a while, which is in line with the assumption of partial asynchronism used in related studies\cite{bertsekas1989parallel}. Region $k$ is allowed to update its local variables after it receives new information from at least $\ceil{p|\mathcal{N}_{k}|}$ neighbors with $0<p\leq1$ and $|\mathcal{N}_{k}|$ denoting the total number of neighbors of region $k$. At worst, any region should wait for at least one neighbor because otherwise its local update makes no progress without any new information. Figure 
\ref{fig:asyncillustration} illustrates the proposed asynchronous scheme by assuming three regions each connecting to the other two regions. The blue bar denotes the local computation and the grey line denotes the message passing. As shown in Fig.
\ref{fig:illusync}, $p = 1$ approximates the synchronous ADMM algorithm, i.e., Algorithm \ref{algADMMsync}, where each region only performs local computation after all neighbors arrive. Figure \ref{fig:illuasync} shows an asynchronous case where each region can perform local update only with one neighbor arrived, which could reduce the waiting/idle time for some regions.

Algorithm \ref{algADMMasync} presents the asynchronous ADMM approach from each region's perspective with $\nu_{k}$ denoting the local iteration counter. The proposed approach does not require any centralized coordination and is applicable to a partial mesh network where each region only communicates with its neighbors. During the initialization, each region solves its local OPF problem without considering its coupling with neighboring regions. In the $z$-update, only the entries in $z_{k}$ associated with the arrived neighbors are updated while the entries associated with the unarrived neighbors remain to be their last updated values. Similarly, $\rho_{k}$ is updated by considering the most recent $\rho_{l}$ received from any neighbor $l$. 

\begin{algorithm}[t]
\caption{Asynchronous ADMM in Region $k$}
\label{algADMMasync}
\begin{algorithmic}[1]
\State \textbf{Initialization} 
\NoNumber{Given ${x}_{k}^{0}$, set ${\lambda}_{k}^{0}={0}$, ${\rho}_{k}=\rho_{0} $, $\nu_{k}=0$ }
\NoNumber{Update ${x}_{k}$ by solving the local OPF}
\begin{align}
\nonumber {x}_{k}^{1} = \underset{{x}_{k}\in \mathcal{X}_{k}}{\text{argmin}} ~f_{k}({x}_{k})\end{align}
\NoNumber{Broadcast $A_{k}x_{k}^{1}$ to region $l$, $\forall l \in \mathcal{N}_{k}$}
\State \textbf{Repeat}
\State\hspace{\algorithmicindent}\textbf{Repeat}
\State\hspace{\algorithmicindent}\hspace{\algorithmicindent}\textbf{Wait} until at least $\ceil{p|\mathcal{N}_{k}|}$ neighbors arrive
\State \hspace{\algorithmicindent}Set $\nu_{k} \leftarrow \nu_{k}+1$
\State \hspace{\algorithmicindent}\textbf{Update}
\NoNumber{~~~~~Update $z_{k}$ associated with arrived neighbors using (\ref{eq:zcalculate})}
\NoNumber{~~~~~Update ${\rho_{k}}$ using (\ref{eq:rhoupdate})}
\NoNumber{~~~~~Update $x$ by solving the local OPF problem}
\begin{align}
\nonumber {x}_{k}^{\nu_{k}} = & \underset{{x}_{k}\in \mathcal{X}_{k}}{\text{argmin}} ~f_{k}({x}_{k})+{\lambda}_{k}^{\nu_{k}-1 \top}(A_{k}{x}_{k}-{z}_{k}^{\nu_{k}})\\
\nonumber &+\frac{{\rho}_k^{\nu_{k}-1}}{2}\|A_{k}{x}_{k}-{z}_{k}^{\nu_{k}}\|^{2}
\end{align}
\NoNumber{~~~~~Update ${\lambda}$ using}
\begin{equation}
\nonumber{\lambda}_{k}^{\nu_{k}}={\lambda}_{k}^{\nu_{k}-1}+({\rho}^{\nu_{k}-1}_k)(A_{k}{x}_{k}^{\nu_{k}}-{z}_{k}^{\nu_{k}})
\end{equation}
\NoNumber{~~~~~Update $\tilde{{\rho}}$ according to (\ref{eq:rhocompute})}
\State\hspace{\algorithmicindent}Broadcast $\{A_{k}x_{k}^{\nu_{k}}, \lambda_{k}^{\nu_{k}}, \tilde{\rho}_{k}^{\nu_{k}}\}$ to region $l$, $\forall l \in \mathcal{N}_{k}$
\State \textbf{Until} a predefined stopping criterion is satisfied
\end{algorithmic}
\end{algorithm}
\color{black} Under the assumption of bounded delay and some other mild conditions on the properties of the considered non-convex problem formulation, the sequence $\{x^{\nu}, z^{\nu}, \lambda^{\nu}\}$ generated by Algorithm \ref{algADMMasync} asymptotically converges to a KKT stationary point $\{x^{*},z^{*}, \lambda^{*}\}$ of problem (\ref{eq:OPF}) with local optimality if a fixed $\rho$ is used. However, same as Algorithm \ref{algADMMsync}, a suboptimal solution might be obtained due to increasing $\rho$. Here, $\nu$ is used to denote the global iteration counter which is increased by 1 whenever a region carries out a local update. We use this global counter only for the purpose of analysis, which is not needed for implementation of Algorithm \ref{algADMMasync}. A rigorous proof for the convergence property of Algorithm \ref{algADMMasync} will appear in a future publication, while we state some of the sufficient conditions here. For the asymptotic convergence of Algorithm \ref{algADMMsync}, $\mathcal{X}$ should be a compact smooth manifold which is indeed the case for the OPF problem, and ${\lambda^{\nu}}$ should be bounded by projecting ${\lambda^{\nu}}$ onto a compact box, i.e., $\lambda^{\nu} \leftarrow \max(\lambda^{\min}, \min(\lambda^{\nu},\lambda^{\max}))$. Furthermore, a local minimum should be identified in the local $x$-update, which is usually observed from our empirical studies and particularly the case when a good partition of the problem is derived such that the coupling among the regions is small \cite{guo2016acase}. 
\color{black}

An important parameter in Algorithm \ref{algADMMasync} is the number of neighbors arrived before next local update. Here $\ceil{p|\mathcal{N}_{k}|}$ only sets the lower bound of the number of neighbors to wait, while in practice, the number of actual arrived neighbors is highly dependent on the communication delays of passing messages. If the communication delay is small compared to the local computation time, it is highly likely that during the time when region $k$ is solving its local problem, the messages from many of its neighbors will arrive. Thereby, the actual arrived neighbors could be much more than the predefined lower bound $\ceil{p|\mathcal{N}_{k}|}$ and region $k$ could immediately start its next iteration without waiting. On the other hand, if the communication delay is large, then region $k$ indeed has to wait even for receiving information from one neighbor. In summary, communication delays determines the severity of asynchronism among the regions, and the larger the delay, the more severe the asynchronism.

Due to the different severity of asynchronism, the penalty ${\rho}$ needs to be carefully chosen to ensure converging to a solution of good quality. The penalty plays a critical role even in Algorithm \ref{algADMMsync}. The larger $\tau$ is, the faster ${\rho}$ increases and consequently the faster ADMM converges, which, however, generally leads to solutions with worse quality since the algorithm proceeds more aggressively to reach any feasible point regardless of its optimality. In the asynchronous case, the penalty $\rho$ needs to be increased at a much slower pace especially under the circumstances where the number of arrived neighbors is small. This is because with partial and delayed information from neighbors, a region tends to make biased decision and therefore needs to proceed with its local iterations with additional caution. 

\section{Simulation Results}
To quantify the convergence performance of any asynchronous distributed approach such as convergence speed and solution quality is generally hard. In this section, we conduct experimental studies to demonstrate the impact of communication delays on the number of arrived neighbors, the choice of penalty parameter and the convergence performance of Algorithm \ref{algADMMasync}.
\label{simulation}
\subsection{Experiment Setup}
The simulations are conducted mainly using the IEEE 118-bus test system. This system is partitioned into eight regions using the partitioning approach proposed in \cite{guointelligent} that has been shown to improve the performance of the ADMM method \cite{guo2016acase}. For each region determined by this partition, the number of neighbors and the average computation time of solving the local OPF problem at one iteration are shown in Table \ref{table:nbor}.
\begin{table}[b]
\caption{Number of neighbors and local computation time of each region}
\centering
\begin{tabular}{ p{2.5cm}<{\centering}| p{0.37cm}<{\centering}|p{0.37cm}<{\centering}| p{0.37cm}<{\centering}|p{0.37cm}<{\centering}|p{0.37cm}<{\centering}|p{0.37cm}<{\centering}|p{0.37cm}<{\centering}|p{0.37cm}<{\centering} }
\toprule
Region&1&2&3&4&5&6&7&8\\
 \midrule
$|\mathcal{N}_{k}|$&3&2&3&1&4&3&3&1\\
\midrule
Computation time ($s$)&0.31&0.13&0.09&0.12&0.15&0.14&0.13&0.11\\
\bottomrule
\end{tabular}
\label{table:nbor}
\end{table}

Algorithm \ref{algADMMasync} is conducted in Matlab R2016a on a personal computer that emulates the process illustrated in Fig. \ref{fig:asyncillustration}. The initialization of $x$ uses a flat start, and the stopping criterion is that the maximum primal residue and constraint mismatch are both smaller than $10^{-3}$ p.u. The initial penalty $\rho_{0}$ is set to 85000, which works well empirically. We use number of local iterations, the execution time and the gap of the objective function to measure the performance of Algorithm \ref{algADMMasync}. The execution time records the total time Algorithm \ref{algADMMasync} takes until convergence including the computation time, the communication delay, and the waiting time for neighbors. For non-convex problems, there is generally a gap between the objective value achieved by the distributed method and the centralized method. The gap in the objective value measures this relative error (in $\%$) of the objective value achieved by Algorithm \ref{algADMMasync} with respect to the one obtained by a centralized method and a solution can be considered of good quality if this gap is smaller than $1\%$.

\subsection{Impact of Communication Delay}
\begin{table}[b]
\caption{Communication delay and number of arrived neighbors}
\centering
\begin{tabular}{ p{1.2cm}<{\centering}| p{1.4cm}<{\centering}|p{1.2cm}<{\centering}| p{1cm}<{\centering}|p{1cm}<{\centering}|p{1cm}<{\centering}}
\toprule
Case&I&II&III&IV&V\\
 \midrule
Delay ($s$)&0.003-0.005&0.03-0.05&0.3-0.5&0.6-1&1.2-2\\
\midrule
$na_{1}$&3.0&3.0&2.6&2.4&1.8\\
$na_{2}$&1.6&1.5&1.3&1.2&1.0\\
$na_{3}$&1.5&1.4&1.4&1.2&1.0\\
$na_{4}$&1.0&1.0&1.0&1.0&1.0\\
$na_{5}$&3.3&3.2&2.9&2.0&1.1\\
$na_{6}$&2.7&2.6&1.7&1.5&1.2\\
$na_{7}$&2.8&2.7&2.4&1.8&1.1\\
$na_{8}$&1.0&1.0&1.0&1.0&1.0\\
\bottomrule
\end{tabular}
\label{table:nborarrive}
\end{table}
To investigate how communication delay affects the actual number of arrived neighbors in Algorithm \ref{algADMMasync}, in the subsequent simulations, we set $p$ to a very small value such that each region can perform its local update as long as it has one neighbor arrived. We consider a wide range of communication delays, and the delay here refers to the time of the message sent from the source region until it arrives at the destination region. For each pair of neighboring regions, its associated communication delay is randomly generated within the range listed in the second row of Table \ref{table:nborarrive}. In power system applications, communication delay could range from a couple of milliseconds to several seconds depending on the infrastructure and technology used. For example, passing message between two regions could take just a few milliseconds using a direct fiber optical link, but could also take a few seconds if there is no direct link available and the message have to be routed through some regional or even central control centers. 

Table \ref{table:nborarrive} shows the average number of arrived neighbors of region $k$ (denoted by $na_{k}$) before its each local update. We calculate this statistic using the first 20 local iterations for each region because the first few iterations are critical in ADMM that determines the final point it converges to. It is shown in Table \ref{table:nborarrive} that with increasing communication delays, the number of arrived neighbor decreases as expected. With small delays such as in Case I and II, all regions can receive the updated information from the majority of their neighbors, while with large delays such as in Case V, most regions have to wait until one neighbor arrives. This indicates that  larger communication delays lead to more severe asynchronism.

\begin{figure}[t]
\setlength{\abovecaptionskip}{0.2cm} 
\centering
\captionsetup[subfigure]{captionskip= 0 cm}
\subfloat[synchronous ADMM]
{
\label{fig:curveall}
\includegraphics[trim = 0mm 0mm 10mm 0mm, clip=true,width=4cm]{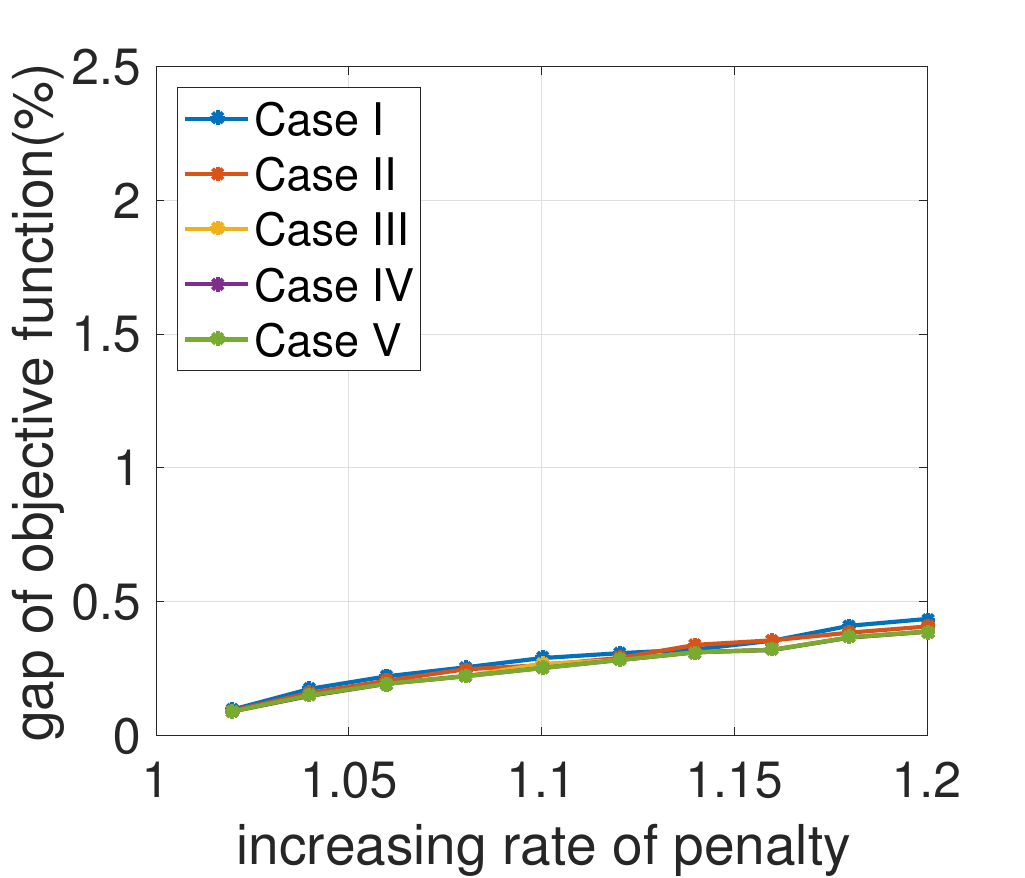} 
}
\hspace{0cm}
\subfloat[asynchronous ADMM]
{
\label{fig:kall}
\includegraphics[trim =0mm 0mm 10mm 0mm, clip=true,width=4cm]{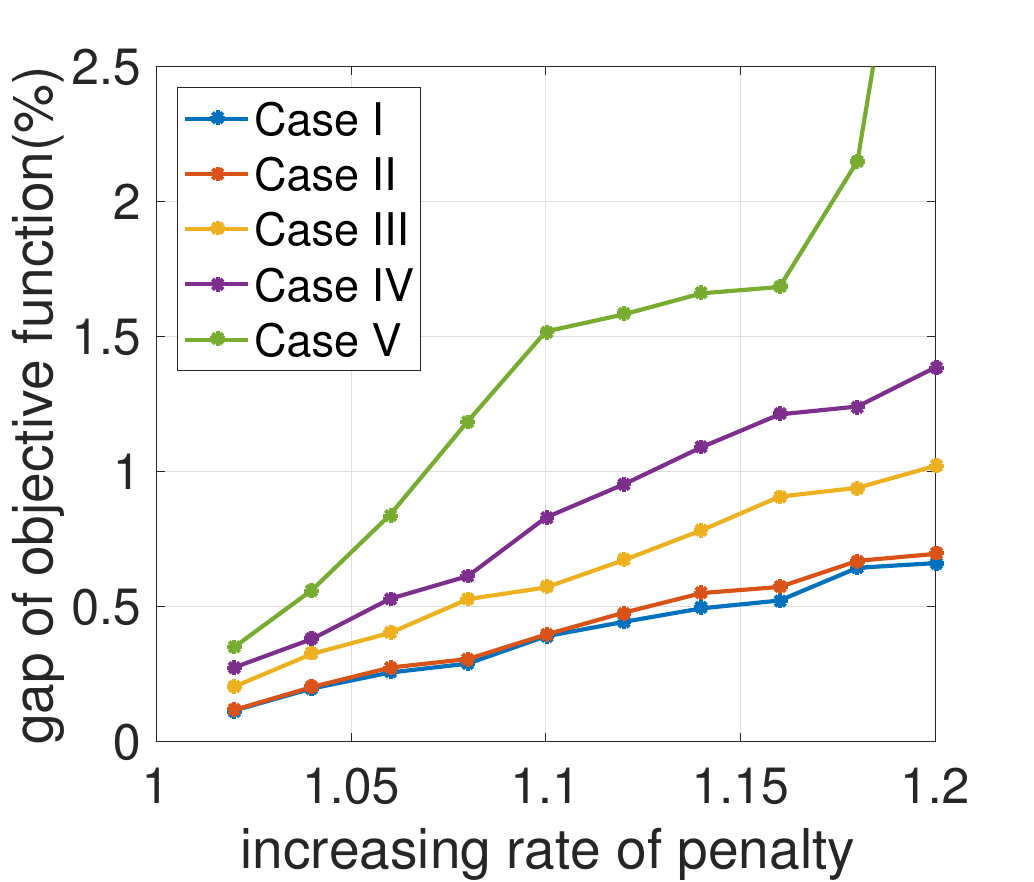} 
}
\caption{Solution quality and the penalty increasing rate.}
\label{fig:gapvstau}
\end{figure}

The severity of asynchronism has a strong impact on the choice of penalty parameter to achieve a solution of good quality. Figure \ref{fig:gapvstau} shows the gap of the objective function between Algorithm \ref{algADMMasync} and the centralized solution with respect to the increasing rate $\tau$ of the penalty $\rho$. For comparison, we also simulated the synchronous counterpart of Algorithm \ref{algADMMasync} where $p$ is set to 1. In the synchronous case, the faster the penalty increases, the larger the gap is. But as all regions need to wait for all neighbors regardless of the communication delay, similar trends can be observed for all cases with various communication delays. In contrast, in the asynchronous case, the sensitivity of Algorithm \ref{algADMMasync} to the increasing rate of penalty highly depends on the communication delays. For example, with large communication delays such as in Case V, one needs to increase the penalty at a very slow pace to reach a solution with a gap smaller than $1\%$.

Table \ref{table:asyncvssync} demonstrates the performance of Algorithm \ref{algADMMasync}  and its synchronous counterpart, i.e., Algorithm \ref{algADMMasync} with $p=1$, with $\nu_{k}^{\max}$, $\nu_{k}^{\min}$ and $\bar{\nu}_{k}$ denoting the maximum, minimum and average local iterations among all regions. The following behaviors of Algorithm \ref{algADMMasync} can be observed: 1) With the same penalty increasing rate, asynchronous ADMM on average takes more iterations than synchronous ADMM but with less time spent on each iteration. Therefore, asynchronous ADMM generally takes shorter time to converge but the solution quality is worse. 2) To achieve similar level of solution quality, asynchronous ADMM needs to adopt a smaller penalty increasing rate that leads to more iterations and possibly longer execution time. 3) With the increase of communication delays, the number of iterations of asynchronous ADMM increases even with fixed penalty increasing rate due to the more severe asynchronism among regions. 4) On this test system, asynchronous ADMM achieves comparable performance of its synchronous counterpart for Cases I to IV where the delay is small or comparable with the computation time, but slows down substantially if the delay is dominant such as in Case V.
\begin{table}[tb]
\caption{Performances of synchronous and asynchronous ADMM under various communication delays}
\vspace{0cm}
\centering
\begin{tabular}{ p{0.6cm}<{\centering}| p{0.8cm}<{\centering}|  p{0.4cm}<{\centering} | p{0.5cm}<{\centering}| p{0.5cm}<{\centering}|p{0.5cm}<{\centering}|p{0.8cm}<{\centering}|p{0.8cm}<{\centering} }
\toprule
Case&Method&$\tau$&$\nu_{k}^{\max}$&$\nu_{k}^{\min}$&$\bar{\nu}_{k}$&Gap(\%)&Time($s$)\\
\midrule
\multirow{ 2}{*}{I}&sync&1.18&36&34&35&0.40&9.7\\
&async&1.18&92&26&56&0.60&7.8\\
&async&1.10&150&45&91&0.40&11.5\\
\midrule
\multirow{ 2}{*}{II}&sync&1.20&31&30&31&0.38&12.3\\
&async&1.20&107&28&61&0.70&10.6\\
&async&1.10&157&48&96&0.40&16.0\\
\midrule
\multirow{ 2}{*}{III}&sync&1.20&48&48&48&0.39&28.8\\
&async&1.20&150&38&89&1.02&14.0\\
&async&1.06&368&119&233&0.40&33.6\\
\midrule
\multirow{ 2}{*}{IV}&sync&1.20&48&48&48&0.39&52.1\\
&async&1.20&222&55&126&1.39&20.9\\
&async&1.04&618&172&377&0.40&57.1\\
\midrule
\multirow{ 2}{*}{V}&sync&1.16&57&56&57&0.32&118.0\\
&async&1.16&390&99&238&1.52&40.0\\
&async&1.02&1735&475&1119&0.35&169.4\\
\bottomrule
\end{tabular}
\label{table:asyncvssync}
\end{table}

\subsection{Application to a Large-Scale Power System}
\label{polish}
To demonstrate the capability of the proposed asynchronous ADMM method on large-scale systems, we apply it to solving the AC OPF problem on the Polish 2383-bus system which is partitioned into 40 regions. The local computation time for each region ranges from 0.05 to 1.2 seconds. The communication delay for each link is within the range of 0.3 to 1 second. As shown in Table \ref{table:asyncPolish}, under considered communication delays, asynchronous ADMM outperforms its synchronous counterpart. The asynchronous scheme is more beneficial for this large system because the control regions in this system are more unbalanced and the synchronous scheme wastes lots of time waiting for slow regions, which is generally the case in large-scale systems.
\begin{table}[tb]
\caption{Performances of synchronous and asynchronous ADMM on the Polish system}
\vspace{0cm}
\centering
\begin{tabular}{ p{0.8cm}<{\centering}|  p{0.4cm}<{\centering} | p{0.5cm}<{\centering}| p{0.5cm}<{\centering}|p{0.5cm}<{\centering}|p{0.8cm}<{\centering}|p{0.8cm}<{\centering} }
\toprule
Method&$\tau$&$\nu_{k}^{\max}$&$\nu_{k}^{\min}$&$\bar{\nu}_{k}$&Gap(\%)&Time($s$)\\
\midrule
sync&1.10&49&46&47&0.54&104\\
async&1.01&2284&37&513&0.43&51\\
\bottomrule
\end{tabular}
\label{table:asyncPolish}
\end{table}
    
\section{Concluding Remarks}
\label{conclusion}
This paper proposes an asynchronous distributed optimization method based on ADMM that allows the control regions in the power system to perform local updates with information received from a subset of its directly connected neighbors. Experimental results show that communication delay significantly affects the number of arrived neighbors, and thereby affects the performance of the proposed asynchronous scheme. Under the settings where communication delay is smaller or comparable with the local computation time, the proposed asynchronous ADMM can achieve comparable or even better performance compared with its synchronous counterpart. These findings indicate that asynchronous distributed methods could be beneficial for large-scale power system optimization but have to be deployed with careful design of the communication infrastructure and schemes. In the future, we plan to investigate other factors that affect the performance of the asynchronous scheme and its applicability to large-scale real-world systems.

\section*{Acknowledgment}

The authors would like to thank ABB for the financial support and Dr. Xiaoming Feng for his invaluable inputs.

\bibliographystyle{IEEEtran}
\bibliography{Partition,Partition2,CommunicationinSG,Asynchronous}

\end{document}